\documentclass[aps,prb,twocolumn,groupedaddress]{revtex4-1}

\usepackage{amsmath}
\usepackage{graphicx}
\usepackage[caption=false]{subfig}

\usepackage{nicefrac}

\usepackage{xspace} 
\newcommand{\msr}{$\mu$SR\xspace}
\newcommand{\fmuf}{F-$\mu^+$-F\xspace}

\newcommand{\F}{Fig.~}
\newcommand{\Eq}{Eq.~}

\newcommand{\yf}{YF$_{3}$\xspace}

\begin{document}


\title{\emph{Ab initio} strategy for muon site assignment in wide band gap fluorides}


\author{F. Bernardini,$^1$  P. Bonf\`a,$^2$, S. Massidda,$^1$ and R. De Renzi$^2$}
\affiliation{$^1$ CNR-IOM and Dipartimento di Fisica, Universit\`a di Cagliari, IT-09042 Monserrato, Italy \\
                   $^2$ Dipartimento di Fisica and Unit\`a CNISM di Parma, Universit\`a di Parma, I-43124 Parma, Italy}


\date{\today}

\begin{abstract}
We report on an \emph{ab initio} strategy based on Density Functional Theory to identify the muon sites. Two issues must be carefully addressed, muon delocalization about candidate interstitial sites and local structural relaxation of the atomic positions due to $\mu^{+}$-sample interaction. Here, we report on the validation of our strategy on two wide band gap materials, LiF and \yf ,  where localization issues are important because of the interplay between muon localization and lattice relaxation. 
\end{abstract}


\maketitle

\section{Introduction\label{sct:intro}}

When collected into a high-intensity spin polarized beam, muons become a powerful probe for many fields of physics and other scientific areas.\cite{yaouanc2010,Cox2009,walker1983,semic,chem} 
Many appealing characteristics have determined the success of muon spin rotation/relaxation spectroscopy (\msr). Firstly,
 \msr significantly widens the classes of materials that can be studied if compared to other spectroscopic techniques (like NMR and ESR for example) since it can be applied to virtually any specimen by simply implanting muons. 
Secondly, during its lifetime the muon spin interacts with magnetic orders of either nuclear or electronic origin, and  provides information on local magnetic fields on a small length scale and - when fluctuating regimes are involved -  on a large frequency window.\cite{yaouanc2010,Stoykov20127} 
Thus, muons are mainly used as a microscopic magnetometer to probe both static and dynamic magnetic orders. Relevant results have also been obtained when modeling the effects of hydrogen-like impurities in semiconductors, and their reaction kinetics, in order to study quantum diffusion.\cite{Cox1987,jpcm-musr2004,Roduner200183}
Moreover, in spatially inhomogeneous systems \msr gives valuable complementary results with respects to neutron diffraction. This also happens with NMR since implanted muons probe the sample from the interstitial region far from nuclei. 
Nonetheless, the muon localization process and the site assignment remain longstanding problems and may represent a serious issue in many \msr experiments since the muon position is often needed to extract quantitative information from $\mu$SR-data. 
After implantation the positive muons ($\mu^+$) usually stop at high symmetry interstitial sites of the crystal lattice. In metals, a cloud of conduction electrons efficiently screens positive charges, 
so that muons leave nearly unperturbed the positions of neighboring atoms.\cite{Hartmann1989}
It is known instead that, due to the formation of chemical bonding between the muon and its neighboring atoms, the final site is in off-center interstitial positions in insulators and semiconductors.\cite{Brewer1986,Brewer1973,Lancaster2007,Hartmann1989,Walle1990} 

From the very beginning of the rise of \msr following the availability of the first experimental facilities, a lot of work has been devoted to the determination of the muon sites. 
A precise characterization of the muon interstitial sites was indeed possible from accurate experimental studies of the Knight shift, of the level crossing resonance (LCR) and by inspecting asymmetry relaxation rates as a function of applied fields in selected compounds.\cite{Camani1977,DeRenzi1984,Chow1994,brewer1991site,Kiefl1988}
Nonetheless, in a large number of cases the muon position and its effect on the hosting system cannot be inferred solely by experimental knowledge, and a reliable method to obtain the muon site in condensed matter would be of great value.
To this aim, a variety of theoretical and computational approaches were used.\cite{Estle1987,Herrero2007,Valladares19951,Cammarere2000,Silva2012,Scheicher2003,Manninen1982,Walle1990,Porter1999}
Successful results have been recently obtained for metallic compounds where an estimation based on the electrostatic potential allowed to identify the muon sites.\cite{Maeter2009}

In this work we want to show that, among the many possible approaches\cite{Maeter2009,Silva2012,Cammarere2000,Kerridge2004} based on a first-principles method, Density Functional Theory (DFT), already well known for its success in studying electronic structure of solids, is further a powerful, accurate and effective tool to explore  the $\mu^{+}$-sample interactions on a selected set of experimental acquisitions. Our work highlights that in insulators the use of a DFT approach is preferable since the stronger interaction with the local environment makes muon position identification a non-straightforward task.

\begin{figure*}
    \includegraphics[scale=1]{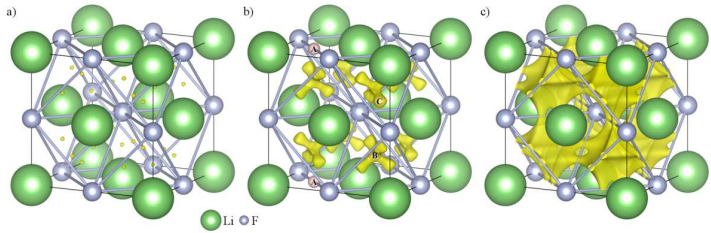}       
    \caption{Isosurfaces of the electrostatic potential in LiF for $V_\mu(r) =$ 50meV, 100meV and 500 meV in a), b) and c) respectively. The isosurface in c) represents the \emph{localization volume} (see text) for the muon in the bulk electrostatic potential.}\label{fig:lifpot}
\end{figure*}

Here we present our results for two fluorine compounds, namely LiF and \yf. They are very useful test cases for our computational investigation. Both materials' ground state electronic structure is well reproduced by DFT. Therefore 
we expect it to provide an accurate value for the electrostatic potential and the local atomic structure surrounding the $\mu^{+}$. 
Among the insulators, LiF and \yf are two well studied cases where a strong $\mu^{+}$-lattice interaction develops. This leads to the formation of a trimer structure with an ionized $\mu^{+}$ between two F nuclei, known as \fmuf complex.\cite{Brewer1986}  The dipolar interaction between F and $\mu$ spins in \fmuf produces a signature in the \msr signal. This allows an accurate determination of the $\mu^{+}$-F distance, which we use to validate our calculations' results.\cite{Lancaster2007}  

The paper is organized as follows. After presenting
the computational strategy for the identification of $\mu^+$ sites in Section \ref{sec:method} we describe the procedure based on the solution of the Hamiltonian for the $\mu^{+}$-F nuclear spin interactions allowing the determination of the local atomic environment from $\mu$SR spectra in Section \ref{sec:experiment}. In Sections \ref{sec:lif} and \ref{sec:yf3} we discuss the outcome of our calculations on LiF and \yf respectively. Finally we draw a summary and conclusion of our work in Section \ref{sec:conclusion}.

\section{Computational details} \label{sec:method}

We use the Generalized Gradient Approximation
(GGA)\cite{LSDA92} to the DFT and the pseudopotential based plane wave method (PPPW) as implemented in the
QuantumEspresso package.\cite{QE} Li, Y and F are somewhat difficult elements for different reasons. F is a first row element with a deep potential, Li and Y have shallow \emph{s} semi-core 
levels. To insure convergence we used Projector Augmented Wave approach\cite{PAW} with explicit treatment of the \emph{s} semi-core as valence and a plane wave basis set up to 800 eV.
The Brillouin zone integration is not a critical issue in those wide band gap insulators, a gaussian smearing with a $8\times8\times8$ Monkhorst-Pack mesh for the bulk and $4\times4\times4$ 
for the supercell ensures convergence. Since we are interested in the electrostatic potential generated by the electron and nuclei our calculations may be in principles biased by the pseudization of the potential inside the atomic core region. For this reason we double checked our muon distribution calculation by comparing the results we got for the perfect bulk system with full-potential calculations using the Augmented Plane Wave plus Local Orbitals (APW+lo) 
\cite{LocalOrbitals,LocalOrbitals2}
method as implemented in the Wien2K package.\cite{wien2k} We found that the muon position is not affected by the approximation in the description of the electrostatic potential in the interstitial region due to the pseudization.

The muon stopping site search problem can be approached as the solution for the motion of a particle slowing down in an effective potential given by a mean field approximation. Here coherently with previous approaches\cite{Maeter2009,Bendele2012} we consider that the potential felt by the muon $V_\mu$ is the sum of the Hartree and nuclei
terms:
\begin{equation}
V_{\mu}({\bf r}) = -\frac{e^2}{2} \int \frac{n({\bf r}')}{{\bf r}-{\bf r}'} dr' + \sum_i \frac{Z_i e^2}{{\bf r}-{\bf R}_i}  
\end{equation}
We call this \emph{the electrostatic approximation} and we disregard possible electron-muon correlation effects. 
Such a potential has minima $V_0$ in the interstitial positions. 
Since muons are light-mass particles it is not trivial the identification of the minima with stopping sites. 
Indeed zero point motion (ZPM) effects may play an important role in muon localization. LiF and \yf are good examples in this respect. 
Because of the ZPM not all of the minima are stopping sites. If we have more than one minimum inside a primitive cell the muon hops between neighboring sites if the barrier to be overtaken is lower than the ZPM energy. At the end the muon will stop in a minimum surrounded by barriers as high as to make further hops impossible, or it will share more than one neighboring minima positions. In the latter case the calculation of the mass center for the ground state will be necessary to identify the muon position.
Inspection of $V_\mu$ in three dimensional systems to understand muon delocalization can be difficult, we need therefore a
criterion to define the extension of the wavefunction about a minimum. We use the turning point concept saying that a wavefunction spreads over the volume of
space, we call \emph{localization volume}, satisfying the condition $V_{\mu}({\bf r}) < E_0$. $E_0$ is the zero point
motion energy (in short zero point energy, ZPE) as the eigenvalue of the ground state solution of the Schr\"odinger equation for the $\mu^+$ particle:
\begin{equation}
\left[\frac{\hbar^2\nabla^2}{2m_{\mu}} + V_{\mu}({\bf r})\right]\psi_{\mu,i}({\bf r}) =
E_{\mu,i}\psi_{\mu,i}({\bf r}).
\label{muschroedinger}
\end{equation}     
As a first approximation, as we did in a recent work
\cite{DeRenzi2012}, ZPEs can be computed by modeling   
each minimum as an anisotropic harmonic well 
\begin{equation}
V({\bf r}) = \frac{1}{2}m_\mu\left(\omega_x^2 x^2 + \omega_y^2 y^2 + \omega_z^2 z^2
\right) + V_0
\end{equation} 
with eigenvalues given by 
\begin{align*}
E(n_x,n_y,n_z) &= \hbar \left[\omega_x(n_x +1/2) + \right. \\
&\left. + \omega_y(n_y +1/2) + \omega_z(n_z + 1/2) \right] + V_0,  
\end{align*} 
and a ZPE
\begin{equation}
E_0=\hbar(\omega_x+\omega_y+\omega_z)/2 + V_0.
\end{equation}  
This method cannot be used whenever the potential well surrounding the minimum has an irregular shape. This is indeed the case here as sections \ref{sec:lif} and \ref{sec:yf3} will show. 
Using this procedure we will end up with disconnected localization volumes each one corresponding to a given stopping site. 
It may also happens that a localization volume goes across the primitive cell boundary connecting neighboring cells. This would correspond to muon diffusion across the crystal, a possibility  
that, as we will show, is prevented by the formation of the \fmuf complex.

If the muon did not modify its environment, that procedure would be sufficiently accurate. 
Instead muons induce  a relaxation of the local neighboring lattice structure.  
Computing the local lattice relaxation we neglect the spread of the muon wavefunction due to the ZPM, then, in the framework of DFT, the effect of muon trapping on the surrounding structure can be studied as it was the trapping process for an
interstitial hydrogen. We set up a supercell up from our bulk crystal and insert an hydrogen
interstitial atom. Since we don't want to reproduce the actual dynamics of the
implantation process for the muons we use an heuristic approach to find the stopping site. We place the
hydrogen impurity in the sites identified as minima by the electrostatic
potential landscape technique or by insights for experiments. Then we let the
system to evolve to the ground state allowing both electron rearrangement and
lattice distortion. The final optimized position for the impurity represents the
refined muon position. Since this approach needs an educated guess
on the position of the muon it should not be considered as totally alternative
to the first. Moreover it does not include any effect due to ZPM.  

The muon is represented by the hydrogen pseudo potential in the
PAW formalism.\cite{PAW} We built our 
supercell up from our bulk structure by doubling the bulk primitive cell along each crystal
axis direction (2$\times$2$\times$2 supercell). 
A convergence threshold of 5 meV is
set for the total energy convergence for structural minimization.
Here we want to study the localization of a $\mu^+$ therefore we make use of charged supercells.
Since charged supercells cannot be treated in the periodic boundary
conditions we use a neutralizing compensating background approach. 
The accuracy of supercell calculations is limited by the size of the simulation.
For neutral light impurities 2$\times$2$\times$2 supercells may be enough. Since we
deal with charged impurities (muon interstitial) we made a convergence test
using SIESTA code.\cite{siesta} Comparing the structure and total energy of
2$\times$2$\times$2 and 3$\times$3$\times$3 supercells we estimate the numerical
error on the energy and on the optimized distances to be $\sim 5$~meV and $\sim 0.02$~{\AA} respectively.

\begin{figure*}
    \includegraphics{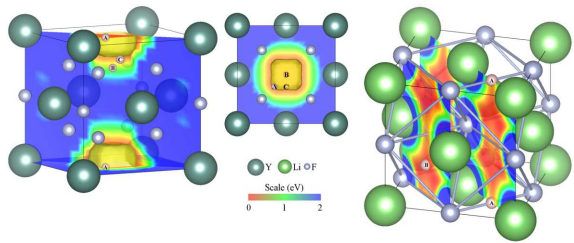}  
    \caption{Possible muon sites in \yf (left and center) and LiF (right). The label $A$ identifies the expected site in both compounds. \emph{Localization volume} surfaces are shown in dark yellow for \yf and in \F\ref{fig:lifpot}(c) for LiF. The electrostatic potential section in \yf and LiF allow a direct comparison the the ZPEs in the two compounds. For sake of clarity, unrelaxed lattice structures are shown.}\label{fig:yf3lif}
\end{figure*}

\section{Solution of the spin-Hamiltonian} \label{sec:experiment}
\begin{figure}
    \includegraphics{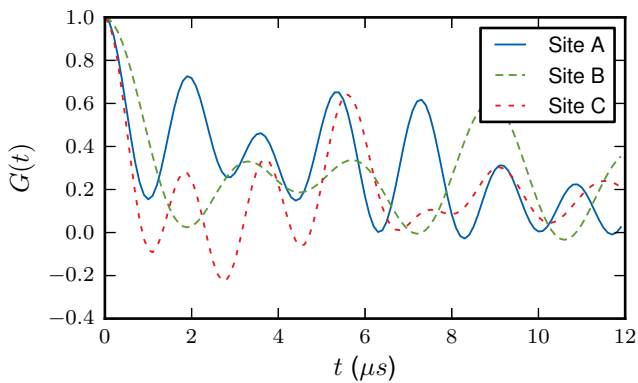}
    \caption{Expected  $\mu^{+}$ asymmetry spectra for optimized muon sites and atomic coordinates in powdered LiF. Sites are indicated as shown in \F\ref{fig:lifpot}. All calculations include all the neighboring atoms giving rise to couplings higher than one tenth of the maximum coupling constant (see Eq.~\ref{eq:hamiltonian}, the number of F atoms considered depends on the $\mu^{+}$ interstitial site). Position $A$ gives the best agreement with the measured data.\cite{Brewer1986}}\label{fig:fmuf}
\end{figure}

LiF and \yf, are especially useful as test cases for our DFT calculations because a precise verification of the muon site is obtained by best fitting the asymmetry signal produced by the dipolar interaction between the muon and neighboring nuclear moments. In fluorides, because of the high nuclear moment of F nuclei (${}^{19}$F has spin $I={}^{1}/_{2}$ and $\sim 100 \%$ natural abundance) and of the high electronegativity of this element, the interaction, commonly referred as \fmuf, is more pronounced. An entangled quantum state develops between the muon and the surrounding nuclei 
and the system may be described with the following Hamiltonian 

\begin{equation}
\label{eq:hamiltonian}
\mathcal{H} = \sum_{i > j} \frac{\mu_{0} \gamma_{i} \gamma_{j}}{4 \pi
  r^{3}} 
\left[ \mathbf{S}_{i} \cdot \mathbf{S}_{j} - 
3 (\mathbf{S}_{i} \cdot \hat{\mathbf{r}})(\mathbf{S}_{j} \cdot \hat{\mathbf{r}}  ) \right], 
\end{equation}
where $\mathbf{r}$ is the vector between spins $S_{i}$ and $S_{j}$ of either the fluorine nuclei or the muon, which have
gyromagnetic ratios $\gamma_{i}$ and $\gamma_{j}$. 
The muon depolarization is given by:

\begin{equation}
    G_{\zeta}(t)= 
    \frac 1 N \sum_{m,n} e^{i(\omega_m-\omega_n)t}\, |\langle m|\sigma_\zeta|n\rangle|^2
\end{equation}
where $N$ is the Hilbert space dimension, $|m \rangle$ and $|n \rangle$ are eigenstates of 
$ \mathcal{H}$ and $\hbar \omega_{m,n}$ are the corresponding eigenvalues, $\sigma_{\zeta}$ is the Pauli spin matrix corresponding to the quantization direction and $h$ is the Planck constant. 
In a powdered sample with cubic symmetry, the observed signal is the results of the weighted average over all directions, i.e.
\begin{equation}
\begin{aligned}
    \overline{|\langle m|\sigma_\zeta|n\rangle|^2} &= \frac 1 3 \left(|\langle m|\sigma_z|n\rangle|^2 + \right. \\
    &+ \left. |\langle m|\sigma_y|n\rangle|^2 + |\langle m|\sigma_x|n\rangle|^2  \right)    
\end{aligned}
\end{equation}

Since the dipolar interaction is inversely proportional to the cube of the inter-nuclear distance, 
one usually consider only up to next neighboring atoms in order to make the calculation of the muon polarization computationally inexpensive within a negligible loss of accuracy. Moreover, the coupling between F nuclear spins may be often disregarded with a limited loss of accuracy even if its inclusion does not lead to an increase of the computational load. 

For an axially symmetric \fmuf complex, as in the case of LiF, when considering only two nearest neighboring F atoms,  the analytic solution of Eq.~\ref{eq:hamiltonian} for a powder averaged depolarization is:

\begin{equation}
  \begin{aligned}
    G_{p}(t) = \frac{1}{6} & \left( 3 + \cos \sqrt{3} \omega_d t + (1-\frac{1}{\sqrt 3})\cos \frac{3-\sqrt 3}{2} \omega_d t \right. \\
     &+ \left. (1+\frac{1}{\sqrt 3})\cos \frac{3+\sqrt{3}}{2} \omega_d t \right) 
  \end{aligned}\label{eq:axfmuf}
\end{equation}
where $\omega_d=\mu_0\gamma_F\gamma_\mu h/(2 r^3)$.
As will be shown hereafter, \Eq\ref{eq:axfmuf} fails to capture the data trend in \yf, and the \fmuf effect alone is not sufficient to determine the $\mu^{+}$ site.

\section{$\bf LiF$}\label{sec:lif}


LiF has the NaCl crystal structure, with a four formula unit conventional cubic cell, containing eight cubic cages with vertexes at four Li and four F atoms. 
As shown in \F\ref{fig:lifpot} the minima of the electrostatic potential in LiF are located approximately at the center of each cage. Minima inside the cage are five. The one we label B in \F\ref{fig:lifpot}(b) is at the very center of the cage surrounded by four equivalent minima labeled C placed in the direction of neighboring F atoms. The minima become connected for $E \geq 75$~meV forming a sort of tetrahedron shaped structure with centroid in site $B$. All of these positions are incompatible with the experimental muon site which is known from literature and was obtained with the strategy explained in Sec \ref{sec:experiment}.\cite{Brewer1986}
\F\ref{fig:lifpot}(c) shows the $\mu^{+}$ localization volume according to the ZPE obtained by the solution of the Schr\"odinger equation for the muon in the bulk electrostatic potential for LiF.
We see that the $\mu^{+}$ is quite delocalized with the localization volume forming a connected network across the crystal. The experimental position, labeled $A$ in \F\ref{fig:lifpot}(b), is at the boundary of the localization volume. 

Muon delocalization inside a strongly polar solid is the condition under which we expect to have a strong effect of $\mu^{+}$-sample interaction on the outcome of our calculations.   
Indeed allowing atoms relaxation in the minimum energy configuration we obtain some large atomic displacement from periodic bulk for all of the sites considered here. 
A strong modification of the crystal structure is found when the muon is added to the interstitial site $A$. While F nuclei are attracted by the charged impurity, Li atoms are repelled. The distance between the muon and its neighboring F nuclei is 1.15~{\AA} in excellent agreement with the experimental data. Also next neighboring F atoms are affected by the $\mu^{+}$ and are subject to a displacement of 0.04~\AA. The relaxed atomic positions correctly describe the formation of a F-$\mu^+$ bonding. 
The relaxed structure for the muon sitting in site $B$ shows a similar behavior: the distance between the muon and its neighboring F atoms reduces from 1.76 {\AA} to 1.57 \AA. We note anyway, that the F-$\mu^{+}$ distance in this case is too large to reproduce the experimental fast decay of the \msr signal.
Site $C$ constitutes a local minimum for the structural relaxation and the hydrogenoid impurity remains trapped there. Anyway the effect of ZPM here is important and 
because of the large delocalization, the muon gets out of the local minimum and reaches site $A$ as a consequence of the gradual atomic positions relaxation.
This behavior is moreover energetically favored if we look at the total energies for the $\mu^{+}$-sample system given by our DFT simulations. 
The total energy for site $A$, $B$ and $C$ are reported in Tab.~\ref{tab:impuritydetails}. We see that the inclusion of relaxation effects allows to recover the agreement with the experimental findings: site $A$ has a total energy which is 0.89 eV lower with respect to site $B$ and is thus confirmed to be the muon stopping site in LiF.
The formation of the \fmuf complex has important consequences on $\mu^+$ delocalization in LiF. Indeed lattice relaxation breaks the lattice periodicity, while the formation of a bond with F enhances the $\mu^+$ localization hindering its diffusion across the material in agreement with the experimental evidence. 

The results of our calculations are confirmed by comparison with experimental data. 
The expected depolarizations for sites $A$, $B$ and $C$ are shown in \F\ref{fig:fmuf}. It is clear that the time dependencies of the muon polarization for the three inequivalent sites are very different allowing us to discard sites $B$ and $C$.
Only site $A$ is compatible with the observed asymmetry spectra, while the other two locations for the $\mu^{+}$ give significantly worse fits (site $C$) and non physical values for the local modification of the bonds length and distances between $\mu^{+}$ and F nuclei (site $B$).  
Fitting the experimental results with $\mathbf{r_{\mu^+-F}}$ as a free parameter in Eq.~\ref{eq:hamiltonian}, we find that the distorted crystal structure obtained from DFT calculations reproduces the experimental F-F distance\cite{Brewer1986} with $\sim 1\%$ precision.
 
 
 \begin{table}
    \begin{tabular}{p{2.5cm}|c|c|c|c|c|c|}
        \cline{2-7}
         \multicolumn{1}{c|}{} & \multicolumn{3}{c|}{LiF}  & \multicolumn{3}{c|}{\yf} \\
         \cline{2-7}
         \multicolumn{1}{c|}{} & $A$ & $B$ & $C$ & $A$ & $B$ & $C$\\
        \cline{1-7}
        \multicolumn{1}{|c|}{F-$\mu^+$ distance [{\AA}]}  & 1.15 & {1.56} &  {1.01} & 0.144 & 1.134 & 1.144\\ \hline 
        \multicolumn{1}{|p{2.5cm}|}{$E_{i}-E_{A}$ [eV]} & 0 & 0.89 & {0.54} & 0 & -0.64 & 0.36 \\ \hline
    \end{tabular}
    \caption{Results for the structural optimization with $\mu^+$ in the interstitial positions A, B and C (see text and \F\ref{fig:yf3lif}). Site $A$ is always the experimental/predicted site. F-$\mu^+$  is the distance between the $\mu^+$ and its nearest neighbor F atom(s), $E_{i}-E_{A}$ is the difference between DFT ground state energies of the relaxed structures.}\label{tab:impuritydetails}
\end{table}

\section{YF$_{\bf 3}$} \label{sec:yf3}  
\begin{figure}
    \includegraphics[]{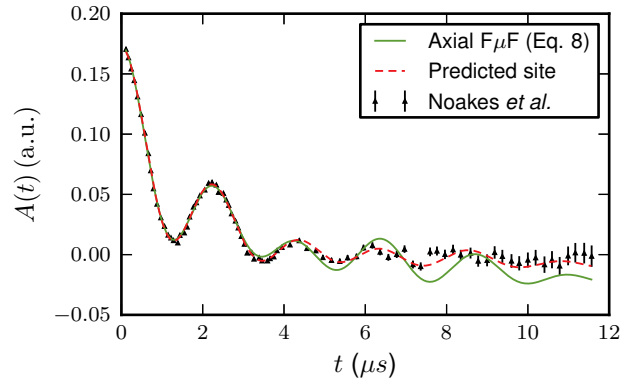} 
    \caption{Fit of \yf data from Ref.~\onlinecite{Noakes1993} with the conventional \fmuf model (\Eq\ref{eq:axfmuf}) and with the depolarization calculated for the DFT predicted site in the fully relaxed structure. The parameters of the fit are detailed in the text and reported in Tab.~\ref{tab:yf3fit}.}\label{fig:yf3fmuffit}
\end{figure}

In order to find the muon sites' positions DFT calculations are more necessary in YF$_{3}$ than in LiF. Firstly because too many inequivalent $\mu^+$ interstitial positions are available in the primitive cell, and so experimental data alone do not allow an unambiguous site identification by the \fmuf signal. Secondly, the Coulomb potential for the unperturbed bulk crystal shows only one minimum in $(\nicefrac{1}{2}, \nicefrac{1}{2}, 0)$ that yields a depolarization which cannot capture the experimental asymmetry spectra.
Moreover here the depolarization signal is only roughly captured by the \emph{axial \fmuf} expectations as shown in \F\ref{fig:yf3fmuffit}.\footnote{It is noted that  authors of Ref.~\onlinecite{Noakes1993} show a better fit to \Eq\ref{eq:axfmuf} of the data with respect to the one in \F\ref{fig:yf3fmuffit}. The difference originates from the assumption of distinct relaxation rates for the constant and the oscillating parts of \Eq\ref{eq:axfmuf}. In this work we compute the expected asymmetry spectra with more atoms then just the first neighboring F nuclei and we also add F-F interactions. As a consequence, the nuclear component of the relaxation rate (for the relevant time interval of the asymmetry spectra) is included in the $G_{F\mu F}$ term of \Eq\ref{eq:yf3fit} and only one phenomenological relaxation rate, of electronic  origin, is introduced in \Eq\ref{eq:yf3fit}.
} 
Therefore in \yf the uncertainty in the muon site assignment can be removed only with the help of DFT calculations. 

Following the same procedure detailed before, we relax the structure with the muon in non-symmetric interstial positions. 
Six possible inequivalent interstitial sites were found after structural relaxations starting from random interstitial positions. 
The three most energetically favorable in-equivalent sites (shown in \F\ref{fig:yf3lif}) are all close to the localization volume. They all are characterized by a slightly distorted \fmuf bond with the muon shifted perpendicular to the F-F axis 
forming, for sites $A$ and $C$, an angle of $\sim 144 ^{\circ}$ between the two bonds. For site $B$ the angle changes to $\sim 160 ^{\circ}$ (in \F\ref{fig:yf3lif} the unrelaxed structures are shown for the sake of clarity). 
The three remaining sites, will not be considered in the rest of the manuscript, since they are too far from the Coulomb potential minimum, have higher ground state energies and result in depolarization functions which are incompatible with the experimental results.
\begin{table}
    \begin{tabular}{lllll}
      \hline
         & $E_{0}$   &  $E_{A}$   &    $E_{B}$   &  $E_{C}$  \\
        \hline 
        LiF   & 0.50          & 0.50          &     0.07     & 0.00   \\
        \yf    &  0.145       &  0.10         &  1.08        &  0.22  \\
        \hline          
    \end{tabular}
    \caption{Zero point energies ($E_{0}$) for the muon in the Coulomb potential minimum and energies for the relaxed muon sites in the bulk Coulomb potential. All energies are in eV and only significant figures are reported.}\label{tab:zeropoinenergy}
\end{table}

\begin{figure}
    \includegraphics[]{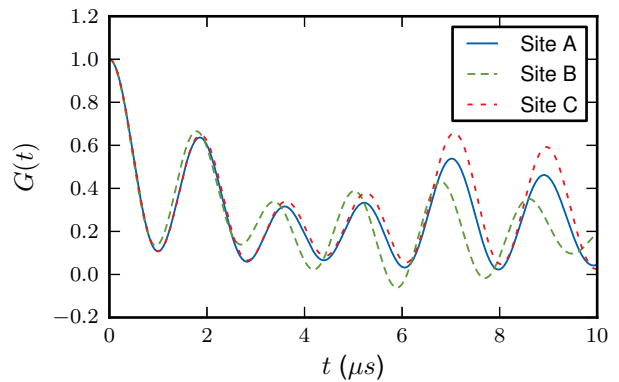} 
    \caption{Expected asymmetry spectra for optimized muon sites and atomic coordinates in powdered \yf. Sites are labeled as in \F\ref{fig:yf3lif}. All calculations include all the neighboring atoms giving rise to couplings higher than one tenth of the maximum coupling constant (see Eq.~\ref{eq:hamiltonian}). Position $A$ gives the best agreement with the measured data.}\label{fig:yf3fmufcomp}
\end{figure}
The depolarizations arising from the relaxed structures of sites $A$, $B$ and $C$ are compared in \F\ref{fig:yf3fmufcomp}. 
The relaxed energies with the relevant parameters obtained from the DFT structural relaxation, are reported in Tab.~\ref{tab:impuritydetails}, while ZPEs and energies for the final muon's positions in the bulk Coulomb potential are given in Tab.~\ref{tab:zeropoinenergy}. 

\begin{figure}
    \includegraphics[]{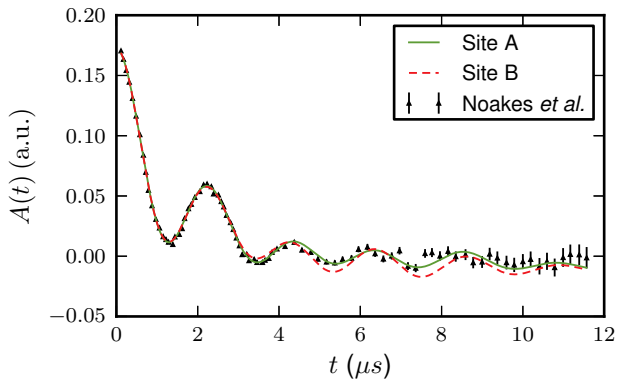} 
    \caption{Fit of \yf data from Ref.~\onlinecite{Noakes1993} with the conventional \fmuf model (\Eq\ref{eq:axfmuf}) and with the depolarization calculated from DFT results obtained for site $A$ and $B$. 
    }\label{fig:yf3fmuffitB}
\end{figure}

In order to identify the muon site, all the above results must be considered.
Indeed, after the structural relaxation, site $B$, which has the lowest energy,  does not provide a correct description of the depolarization function (\F\ref{fig:yf3fmuffitB}). 
Instead a good description of the experimental data is obtained when considering the expected depolarization from site $A$, as shown in \F\ref{fig:yf3fmuffit}. 
Moreover, we point out that kinetics should favor trapping in site $A$, as this is the site with the lowest electrostatic potential in the unperturbed bulk structure and the only one whose position is inside the   $\mu^+$ localization volume of the $V_\mu$ global minimum. Both site $B$ and $C$ have energies higher than the muons's ZPE. 
Thus the probability of finding the muon in site $B$ is lower with respect to sites $C$ and $A$ and therefore the formation of a \fmuf complex in $A$ is more likely. 
This findings led us to conclude that the \fmuf complex in site $A$ is the maximally populated muon site in \yf. 
As for LiF, the localization region shrinks as a consequence of the formation of the bond, allowing us to neglect the ZPM when solving Eq.~\ref{eq:hamiltonian}

\begin{table}
    \begin{tabular}{cccc}\hline
         & Conv. \fmuf & Site $A$ & Site $B$ \\
        \hline 
         $A_{0}$               &  0.202(1)               &  0.184(1)             &  0.188(1)              \\
         $p_{1}$               &  0.77(1)                &  0.76(1)              &  0.75(1)              \\
         $A_{calbg}$           & -0.028(1)               & -0.012(1)             & -0.014(1)              \\
         $\lambda_{F\mu F}$    &  0.18(1) $\mu$s$^{-1}$  &  0.19(1)              &  0.15(1)              \\          
         $\beta$               &  1.27(6)                &  1.45(1)              &  1.23(1)                \\ 
         $r_{F-\mu}$           &  1.23(1) \AA            &  1.17(1)  \AA         &  1.22(1) \AA          \\ 
         $\sigma$              &  0.73(1)  $\mu$s$^{-1}$ &  0.97(1) $\mu$s$^{-1}$&  1.00(2) $\mu$s$^{-1}$ \\          
         $\chi^{2}_{r}$        &  4.7                    &   2.3                 &  4.1 \\
         \hline
    \end{tabular}
    \caption{Parameters for \Eq\ref{eq:yf3fit} obtained from the best-fit to the data of \F\ref{fig:yf3fmuffit}. In the first column the results obtained with $G_{F \mu F}$ defined in Eq.~\ref{eq:axfmuf} (already obtained by the authors of Ref.~\onlinecite{Noakes1993}) are reported. In the second and third columns $G_{F \mu F}$ is calculated from DFT results (see text). $r_{F-\mu}$ is the distance between the muon and the first neighboring F atom(s) for a given $\mu^{+}$ site. 
    Small discrepancies between the experimental and calculated $r_{F-\mu}$ (that may possibly arise from the reduced but non-vanishing ZPM neglected in \Eq\ref{eq:hamiltonian}) are accounted by the $\delta\omega$ parameter (see text). The final F-$\mu$ distance is obtained by conveniently scaling all the distances between the $\mu^{+}$ and the atoms included in the sum of \Eq\ref{eq:hamiltonian}. All the scaling factors are smaller than 5\%. }\label{tab:yf3fit}
\end{table}

The experimental data were fitted according to the equation:
\begin{equation}
  \begin{aligned}
    A(t) &= A_{0} \left[ \right. p_{1} G_{F \mu F}(t, \delta\omega) \exp\left(-(\lambda_{F \mu F} t)^{\beta} \right)  \\
    &+  p_{2} \exp\left(-(\sigma t)^{2}\right)  \left. \right] + A_{calbg} \label{eq:yf3fit}
  \end{aligned}
\end{equation}
 where $A_{0}$ is the total asymmetry arising from the sample and the sample holder, $p_{1}$ measures the fraction of muons reaching the \fmuf site, $p_{2} = 1 - p_{1}$ and $\sigma$ account for the depolarization in the presence of weak nuclear coupling and $A_{calbg}$ is added in order to compensate for the background and for the uncertain calibration of the non precessing component. $G_{F\mu F}$ is obtained by solving \Eq\ref{eq:hamiltonian} with the lattice structure obtained from DFT calculations and $\delta\omega$ is a parameter that accounts for small discrepancies between F-$\mu^+$ calculated and experimental distances.

The parameters obtained from the best fits shown in Figs.~\ref{fig:yf3fmuffit} and \ref{fig:yf3fmuffitB} are reported in Tab.~\ref{tab:yf3fit}. We finally add that site $C$ is also compatible with the experimental data and therefore we cannot rule out the possibility of a partial occupation of this site.

\section{Conclusions}\label{sec:conclusion} 

In this work we show that, in materials where a strong $\mu^+$-system interaction is present, the correct interpretation of $\mu$SR experiments
requires a combination of experimental and theoretical investigation. The latter is best done by an {\it ab initio} approach within DFT. Indeed DFT is able to provide on the same footing the electrostatic potential we use to find candidate sites for $\mu^+$ localization and, by the impurity approach, allows us to get the refined atomic structure we use to interpret the experimental data. 
We tested our procedure in LiF, where we showed that the bulk electrostatic potential fails to correctly predict the actual $\mu^+$ site, while upon structural refinement we were able to reproduce 
the formation of the \fmuf complex and its structural details (F-$\mu^+$ distance). 
We than extended our investigation on \yf where the presence of several candidate interstitial site makes impossible
the identification of the $\mu^+$ position from experimental knowledge alone. Comparing the experimental data with the refined structure obtained by DFT investigation we were able to  predict the correct location 
and shape for the \fmuf complex in \yf. We point out that such an approach, we testes on materials were $\mu^+$-system interaction is quite large, is of general validity and can be applied on a wide choice of different material other than wide band gap insulators.\cite{nota}

\section{Acknowledgments}
P. B. and R. D. R. want to acknowledge S. R. Kreitzman, J. Moeller, D. Ceresoli and T. Lancaster for fruitful discussions. 
F. B. acknowledges support from CASPUR under the Standard HPC Grant 2012 and from the FP7 European project
SUPER-IRON (grant agreement No. 283204).

\bibliography{muon}
\end{document}